\documentclass{aa}
\usepackage{epsfig}

\newcommand{\taud}{$\tau_{\rm disk}$}
\newcommand{\optheta}{$\theta_{\rm op}$}

\newcommand{\rsun}{R$_{\odot}$}
\newcommand{\Rsun}{R$_{\odot}$}
\newcommand{\Rstar}{R$_{*}$}
\newcommand{\rstar}{R$_{*}$}
\newcommand{\Rin}{R$_{\rm in}$}
\newcommand{\rin}{R$_{\rm in}$}

\newcommand{\Msun}{M$_{\odot}$}

\newcommand{\degree}{$^{\rm o}$}

\newcommand{\ha}{H$\alpha$}

\newcommand{\bb}{\bibitem[]{bla}}

\def\lesssim{\mathrel{\hbox{\rlap{\hbox{\lower4pt\hbox{$\sim$}}}\hbox{$<$}}}}
\def\gtrsim{\mathrel{\hbox{\rlap{\hbox{\lower4pt\hbox{$\sim$}}}\hbox{$>$}}}}

\begin{document}


\title{Polarimetric line profiles for scattering off rotating disks}

\author{Jorick S. Vink{$^1$}, Tim J. Harries{$^2$}, Janet E. Drew{$^1$}}
\offprints{Jorick S. Vink, j.vink@imperial.ac.uk}
\institute{$^1$Imperial College London, Blackett Laboratory, Prince Consort Road, London, SW7 2BZ, UK\\   
           $^2$ School of Physics, University of Exeter, Stocker Road, Exeter EX4 4QL, UK}

\titlerunning{Polarimetric line profiles}
\authorrunning{Jorick S. Vink et al.}

\abstract{We predict polarimetric line profiles for scattering off rotating disks using a 
Monte Carlo technique. 
We have discovered that there is a marked difference between scattering of line emission 
by a disk that reaches the stellar surface, and a disk with an inner hole. 
For the case with an inner hole, we find {\it single} position-angle 
rotations, similar to those predicted by analytic models. 
For the case of an undisrupted disk, we find {\it double} rotations in 
the position angle -- an effect not reported before. We show that this new effect 
is due to the finite-sized star interacting with the disk's rotational velocity 
field.
Since a gradual increase of the hole size transforms the double rotations smoothly back 
into single ones -- as the line emission object approaches that of a point source -- our 
models demonstrate the diagnostic potential of line polarimetry in determining not only the disk 
inclination, but also the size of the disk inner hole.
Our models are generic, and relevant to any line emitting object that is embedded in a 
rotating disk. Examples are: Herbig Ae stars, T Tauri stars, 
other young stars, 
early-type stars with disks, post-AGB stars, cataclysmic variables and 
other binary systems, as well as extra-galactic objects, such as the disks around super-massive 
black holes.
}

\maketitle


\section{Introduction}
\label{s_intro}

In this paper, we present line polarimetry predictions for scattering 
off rotating disks using a Monte Carlo technique.
Our models are generic, and relevant to a variety 
of emission line objects incorporating disks. This could include Herbig Ae stars, T Tauri stars, 
other young stars, early-type stars with disks, post-AGB stars, cataclysmic variables and 
other binary systems, as well as extra-galactic objects, such as the disks around super-massive 
black holes.

Linear spectropolarimetry is a powerful tool to study the geometry 
and kinematics of the innermost regions around a variety of 
objects (Drew et al. 2004).
First, the spatial scales on which these environments need 
investigation (such as rotating disks, accretion flows, winds) 
are generally too compact to be directly imaged -- a situation that 
will persist even beyond the commissioning of 100m-class telescopes.
Second, straightforward spectroscopy aiming to deduce geometric and 
kinematic information from spectral line profiles is generally not unique, 
because of the convolution that underlies spectral line 
formation. For instance, the \ha\ spectral line is often used to 
determine mass accretion rates in T Tauri stars and brown dwarfs 
(e.g. Muzerolle et al. 2003), whereas others use it to derive wind mass-loss rate 
in young stellar objects (e.g. Nisini et al. 1995). 
This illustrates the fact that more 
than one process may contribute to the line emission. 
Since linearly polarized light is in most cases scattered light, whilst unpolarized 
light is directly sourced light (e.g. coming directly from the star), 
the advantage of spectropolarimetry is its access to 
an extra dimension of information, 
which can provide vital clues 
to otherwise unresolved geometries. A telling
success of spectropolarimetry was its role in underpinning the `unification'
model of active galactic nuclei (e.g. Antonucci \& Miller 1985). 

In the stellar context, spectropolarimetry has strongly 
motivated the now widely-accepted circumstellar disk model for 
classical Be stars (Wood et al. 1997, Quirrenbach et al. 1997). 
The technique of ``line polarimetry'' -- measuring polarization changes across an individual 
emission line -- has widely been applied to these 
somewhat evolved stars (e.g. Poeckert 1975). 
It has recently also been applied to pre-main sequence (PMS) stars 
(Oudmaijer \& Drew 1999, Vink et al. 2002, 2003). 
For classical Be stars, the dominant observed effect is known to be caused by  
unpolarized line emission in the presence of intrinsic continuum 
polarization (e.g. Clarke \& McLean 1974). This `depolarization' 
across emission lines occurs when line photons are 
formed over a larger volume (e.g. in the circumstellar disk) than 
the continuum photons. The line photons 
therefore scatter less often -- and over a wider range of scattering angles -- 
off the electrons in the disk. As a result, a smooth
change in the polarization 
percentage and/or position angle is seen across the line profile.

There are circumstances where instead  
the {\it line} photons are formed closer to the star, 
and these are scattered and polarized {\it themselves}. Such 
effects have recently been detected in 
both low and intermediate mass T Tauri and Herbig Ae stars 
(Vink et al. 2002, 2003, 2004), where the \ha\ line is thought to be 
polarized by scattering in a rotating accretion disk. Analogous results 
have been obtained from observations of Seyfert 1 galaxies by Smith 
et al. (2002). 

Although the \ha\ spectropolarimetry data themselves are 
powerful enough to prove that the geometries 
around these objects are not spherically symmetric, the observations 
carry additional information on both the photon source and the kinematics 
(such as rotation) of the circumstellar medium. For an optimum use 
of spectropolarimetry data, it is crucial that  
modelling tools are developed that avoid geometric simplifications. 
Such models have -- until now -- not been available. 
First steps were undertaken by Wood, Brown \& Fox (1993; hereafter WBF) 
who analytically studied the polarization and position angle (PA) 
of line photons from a uniformly radiating point source that are scattered 
within a dynamical scattering medium, such as the rotating disks of classical Be stars. 
We update here to a 
Monte Carlo approach that allows more realistic geometrical 
effects, such as the finite-sized photon source, to be explored. 
We will see that finite-star photon emission has subtle 
effects on the line polarization that have gone unnoticed in prior
analytic studies. Note that we do not intend to reproduce the continuum and line polarization
of specific object classes. Instead, we adopt an (idealized) picture
of compact line emission that is scattered in a rotating disk. The surprising transition  
in the line polarization profiles between a finite-sized star and a point source, 
may provide a unique tool for constraining the size of an inner disk hole for a 
variety of centrally condensed objects that are embedded in disks.

The paper is organized as follows. In Sect.~\ref{s_descr}, we discuss 
the basic features of the Monte Carlo model and the disk geometries. 
In Sect.~\ref{s_test}, we test the Monte Carlo model against the analytic 
results of WBF for a point source, before we 
present our Monte Carlo results for a finite-sized photon source in Sect.~\ref{s_line}. 
Finally, we discuss (Sect.~\ref{s_discuss}), and summarize the main outcomes of 
our study (Sect.~\ref{s_sum}).


\section{Description of the geometry and the Monte Carlo model}
\label{s_descr}

The 3D Monte Carlo code {\sc torus} is described in Harries et al. (2000) 
where it is applied to hot star winds. Here, we consider a rotating disk 
geometry. Since we are especially interested in accretion 
disks around PMS stars, where the role of inner disk disruption by magnetic 
fields has gained importance (see e.g. Edwards et al. 1994, Muzerolle et al. 1998), 
we have implemented a geometry in which the inner disk can be truncated. 
This allows us to study the influence of an inner hole on the resulting line polarimetry.

Our study is aimed at understanding different scattering geometries, and the 
radiative transfer is therefore simplified (but see Hillier 1996). 
Following WBF, we begin with 
line emission that is scattered within a rotating disk. 
Since our study is primarily aimed at understanding geometrical effects, we 
consider the radial optical depth in the disk mid-plane \taud, rather than the column density 
of a particular particle type (such as electrons or dust grains).
Also, the scattering matrix we employ is the Rayleigh scattering matrix (correct for  
electron and resonance-line scattering, as well as for small dust grains, but for scattering off larger 
dust particles the Mie phase function would be more appropriate, see also Henney 1994). The focus of this study is on the scattering 
geometry rather than on the identification of the polarizing agent in a particular 
object class -- a task which may be tackled by broader band spectropolarimetry (e.g. Bjorkman et al. 2003) 
rather than by line polarimetry. 
In any case, we do not expect the choice of scattering matrix 
to qualitatively change the differential polarization behaviour across the line.

Most of our Monte Carlo calculations are run with $10^6$ photons, but 
to keep up good statistics for lower inclination (pole-on) disks, this number 
is increased to $10^7$, or more, where necessary.
We compute the linear Stokes parameters $Q$ and $U$, from which 
the polarization percentage ($P$) and the PA ($\theta$) 
can readily be obtained:

\begin{equation}
P~=~\sqrt{(Q^2 + U^2)}
\label{eq_defp}
\end{equation}
\begin{equation}
\theta~=~\frac{1}{2}~\arctan(\frac{U}{Q})
\label{eq_defpa}
\end{equation}
We have adopted a Stokes $I$ line width with a FWHM of 300 km/s. 
The default line/continuum contrast is 3. This combination of line width and contrast 
results in a line profile that is very similar to that in WBF. 
These contrast levels are also typical for \ha\ observed in Herbig Ae 
stars (see for instance Vink et al. 2002).

\subsection{Description of the disk geometry}

The disk geometry is set up using spherical polar coordinates, with 
finer gridding of polar angle within a few degrees of the disk equator. 
We start off with a flat disk geometry, in order to compare our Monte Carlo
models with the analytic thin disk results of WBF. From then onwards, we will switch 
to a disk with a constant opening angle $\theta$ -- the ``theta'' disk. 
The reason for this is to limit the parameter space: disks with a range of different 
inner-hole sizes will subtend similar solid angles, and are hence expected to 
yield similar polarization levels for a larger range of models. Furthermore, 
a theta disk seems more appropriate for an accretion disk than is a flat disk.
We note that the differences between a flat and 
our theta disk are only subtle as far as the predicted 
polarimetry behaviour across lines is concerned. This is because the polarization 
in our models turns out to be 
mostly dependent on the optical depth in the inner disk region.

\subsubsection{The Flat disk}

For the case of a flat disk, the density, $\rho(r)$, of the disk falls off 
exponentially over the disk scale-height $H$, which is held constant in the radial direction. 
In the disk mid-plane, both $\rho$ and $\Sigma$, the surface density, show the following 
behaviour in the radial direction:

\begin{equation}
\rho(r)~\propto~r^{-2}~~~~~~~~\Sigma(r)~\propto~r^{-2}
\label{eq_sig}
\end{equation}
where $\Sigma$ $=$ $2 \rho H$ is the surface density at the 
inner disk radius. 
The radial optical depth in the disk mid-plane is:

\begin{equation}
\tau_{\rm disk}~\propto~\int_{R_{\rm in}}^{R_{\rm out}} \rho(r) dr   
\label{eq_sig}
\end{equation}
where the radially integrated disk optical depth \taud\ is chosen in the range of approximately 
1 to 10. Note that the disk outer boundary is chosen to 
be 50 solar radii, but because the density falls off rapidly, the exact choice 
of outer boundary radius is 
irrelevant. 

The disk inclination to the observer's line-of-sight can be varied from pole-on 
to edge-on, where we define $i$ $=$ 0\degree\ to be pole-on, and $i$ $=$ 90\degree\ to be edge-on.
The Keplerian rotation is anti-clockwise, with a rotational velocity:

\begin{equation}
v_{\phi}(r)~=~\sqrt{(\frac{G M}{r})}
\label{eq_grav}
\end{equation}
where $G$ is the gravitational constant. Equation~(\ref{eq_grav}) implies
that the highest velocities are reached close to the stellar surface. At the 
surface, the rotational velocity is about 440 km/s, since we have adopted 
\Rstar\ $=$ 3\Rsun, and $M$ $=$ 3\Msun, as is typical for Herbig Ae stars (e.g. 
Hillenbrand et al. 1992). Note that we have also explored a solid-body rotational 
velocity field (so that velocity increases with radius), but because the 
line polarimetry is critically dependent on the innermost part of the disk, the 
choice of {\it the exact} character of the velocity field turns out to be a minor factor. 
Although this implies 
that line polarimetry cannot be used to discriminate a Keplerian velocity field 
from solid body rotation, it does mean 
that polarimetric line profiles can be a robust way of determining 
the disk inner edge radius and velocity.

\subsubsection{The theta disk}

If we would consider an accretion disk model more commonly invoked in the discussion 
of accretion young stellar objects (e.g. Carr 1989, Chandler et al. 1995), the disk 
surface density would be proportional to:

\begin{equation}
\Sigma(r)~\propto~T^{-1} r^{-3/2}
\end{equation}
However, since the temperature in both a ``passive'' reprocessing disk, as well 
as an ``active'' accretion disk is proportional to $T \propto r^{-3/4}$ (see respectively 
Kenyon \& Hartmann (1987) and Shakura \& Sunyaev (1973) for details), it would follow 
that the surface density behaves as 

\begin{equation}
\Sigma(r)~\propto~r^{-3/4}
\end{equation}
Since the half-thickness scale-height $h = H/2$ depends quasi-linearly 
on radius, as

\begin{equation}
h(r)~\propto~T^{1/2} r^{3/2} \propto r^{9/8}
\end{equation}
(assuming $T$ to be isothermal perpendicular to the disk mid-plane), 
the density within a theta disk would be 

\begin{equation}
\rho = \frac{\Sigma}{2 h} ~\propto~\frac{r^{-3/4}}{r^{9/8}} ~\propto~r^{-15/8}
\label{eq_carr}
\end{equation}
This radial behaviour of Eq.~(\ref{eq_carr}) is very similar to a density falling off as $\rho$ $\propto$ $r^{-2}$, 
as used in the flat disk case (Eq.~\ref{eq_sig}), and we continue to use it in our description
of the theta disk. Finally, note that the parameter $H$ does not play any role in the 
theta disk description any more. Instead, a half-opening angle \optheta\ for the disk 
is defined, which plays a role analogous to that of $H$ in the flat disk case.

Note that we have used both disk geometries, i.e. the flat and theta disk,
for the tests in Sect.~\ref{s_test} as well as for the results in Sect.~\ref{s_line}, but 
have decided to present only the flat disk in the testing part, and the theta disk 
for our subsequent results thereafter. 


\section{Testing the Monte Carlo disk model}
\label{s_test}

We first test whether a uniform photon source illuminating an almost 
pole-on disk, yields ``zero'' net 
polarization. This is indeed found to be the case. 

\begin{figure}
\centerline{\psfig{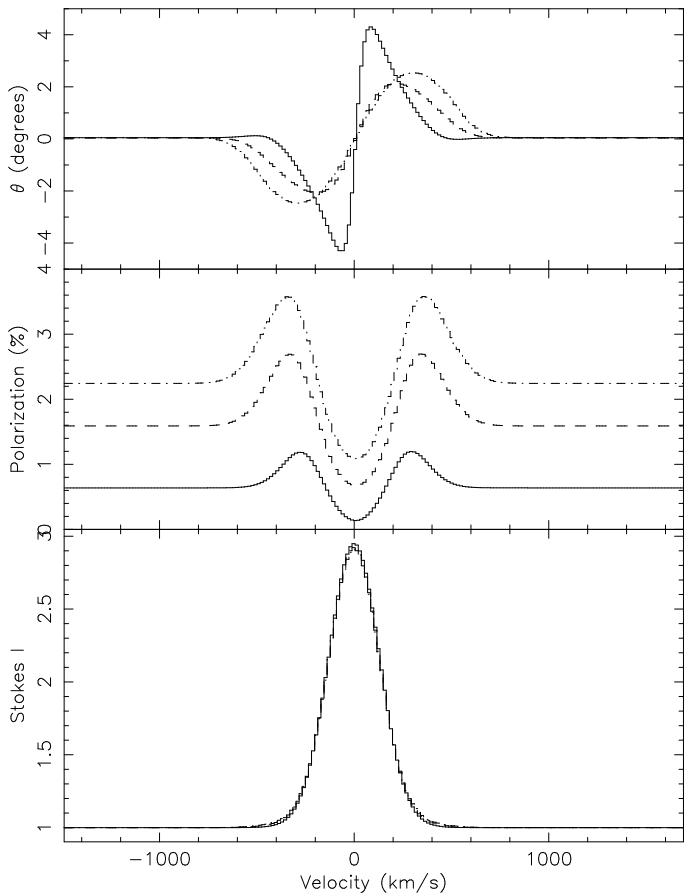}}
\caption{Triplot of the polarization spectra for a rotating disk 
seen at 3 different inclinations. The assumed Stokes I spectrum is shown 
in the lowest panel, the degree of linear polarization \%Pol is 
indicated in the middle panel, whilst the PA ($\theta$; see Eq.~\ref{eq_defpa}) 
is plotted in the upper panel. 
The solid line is for the disk seen at 30\degree, the 
dashed line at 60\degree, and the dashed-solid line for 85\degree, 
which yields the largest continuum polarization, and the smallest 
PA line-centre rotation.}
\label{f_wood}
\end{figure}

As far as continuum polarimetry is concerned, Brown \& McLean (1977) computed 
the continuum polarization of a point source and single electron scattering 
analytically, and found that 

\begin{equation}
P (\%)~\propto~\tau_{\rm mean}~(1 - 3\gamma) \sin^{2}i
\label{eq_brown}
\end{equation}
where $\gamma$ is a number between 0 and 1 that represents their envelope shape factor.
$\gamma$ = 0 gives a flat disk, $\gamma$ = 1/3 yields a spherical envelope (resulting in zero 
polarization), whilst the envelope becomes prolate for $\gamma$ $\ge$ 1/3.
$\tau_{\rm mean}$ is the mean scattering optical depth. This is 
much smaller than \taud, since it represents the typical optical depth between 
the source and the scatterer, whilst our \taud\ is the optical depth integrated across 
the entire disk; Eq.~(\ref{eq_sig}).

We ran a grid of disk models, and studied the continuum polarization. 
Contrary to Brown \& McLean (1977), we use a finite star (see also Cassinelli et al. 1987 and 
Brown et al. 1989) and employ multiple scattering with our Monte Carlo model. 
We varied: (i) the disk inclination $i$, the (ii) disk scale-height $H$, as well as (iii) the 
integrated disk optical depth \taud.
From these Monte Carlo models, we find the continuum polarization to behave 
similarly to that of Brown \& McLean for a limited density regime corresponding 
to \taud\ $\la$ 10. For higher optical depths, multiple scattering effects become 
noticeable, and the degree of polarization no longer increases according to Eq.~(\ref{eq_brown}). 
In addition, the PA starts to behave differently: as long as multiple scattering is unimportant, 
the PA is perpendicular to the plane of scattering (i.e. the disk), but when 
multiple scattering effects become significant, the bulk of effective scattering 
occurs in the vertical disk direction, and ultimately the PA will lie within the 
same plane as the disk (see also Angel 1969, Bastien \& Menard 1988).
 
We now compare the {\it line} polarization of our 
multiple scattering model with the analytical  
results of WBF for a range of viewing angles. 
Since WBF used a single-scattering model with a stellar point 
source and an infinitely thin disk, while we consider a finite-sized star, a 
disk of finite thickness (which is a numerical necessity), and multiple scattering, 
we do not seek for perfect quantitative agreement. However, we require the 
continuum polarization to be of the same order of magnitude. 
Our test calculations are performed by using an inner hole to effectively 
reproduce a point source. This 
was done by decreasing the stellar core size from 3 to 0.3 \rsun,
while the inner disk radius was left at 3 \rsun. 
To stay close to the results of WBF, we chose a geometrically thin disk, with a 
disk scale-height $H$ of 0.2 \rsun. The integrated disk optical thickness \taud\ was 
chosen to be about 10.

Figure~\ref{f_wood} shows line polarimetry results for a disk 
viewed under inclination angles of 30\degree, 60\degree, and 85\degree. 
The overall polarization results are very similar to those in Figure~4 
of WBF. Entirely analogous to WBF the symmetric 
shape of the polarization spectrum is seen in the middle panel of 
Fig.~\ref{f_wood}, with the highest \%Pol for the most inclined 
disk (85\degree), as expected. 
The reason for the stronger polarization in the line
wings and the dip at line centre is due to the fact that the unscattered line 
emission width (with an adopted FWHM of 300 km/s) is 
narrower than that of the scattered light (the disk inner-edge rotation 
velocity is 440 km/s). 
Dividing broad by narrow 
emission yields the characteristic line polarization shapes as seen in 
the middle panel of Fig.~\ref{f_wood}.

The PA, plotted in the top panel, is seen to rotate 
by a few degrees at line centre. Such PA changes are known to transform into `loops' 
when plotted in $QU$ space (see Vink et al. 2002 for cartoons). 
Such $QU$ loops were first discussed by Poeckert \& Marlborough (1977), and 
McLean (1979) in the context of a rotating disk around classical Be stars 
suffering line absorption. However 
WBF and Ignace (2000) have noted similar S-shaped PA rotations 
due to the occulting effect of the star, as 
photons from the dark side of the object are not able to reach the observer. 
Due to the asymmetry in velocity between the red and blue side 
of the line, the crucial front-back asymmetry allows Stokes $U$ to change 
sign, and the PA rotates from negative to positive, causing the line-centre 
flip seen in the upper panel of Fig.~{\ref{f_wood}. 

Interestingly, if we do not take occultation into account, 
but allow the star to be transparent,
the S-shaped position angle rotation still remains. We have tested whether the PA 
rotation is then due to our multiple-scattering approach. Indeed, if we allow photons to 
scatter only once, the S-shaped PA rotation is absent. This shows 
that multiple scattering introduces an analogous front-back asymmetry 
effect to that due to stellar occultation, which is a new result.

\section{Line polarimetry from a finite-sized star}
\label{s_line}

\begin{figure}
\centerline{\psfig{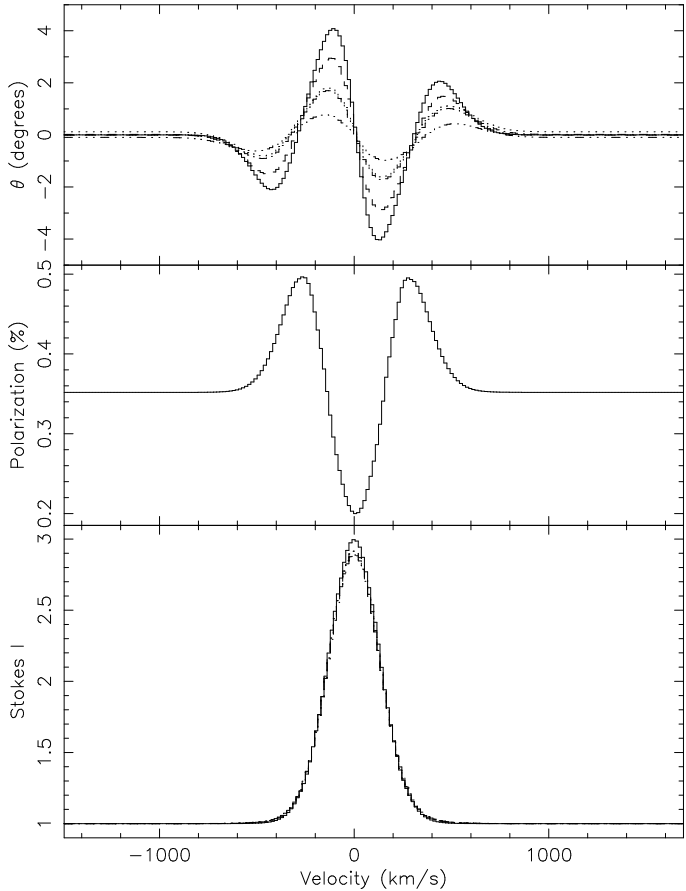}}
\caption{The case of the undisrupted disk. The model inclinations are 15\degree, 30\degree, 
45\degree, 60\degree, and 75\degree, with the lowest inclination 15\degree\ model represented by 
the solid line, and a gradual decrease in the PA amplitude for increasing inclinations.
Note that although the level of continuum polarization 
is higher for higher inclinations -- typically reaching a few percent -- they are not all shown, 
as neither the widths, nor the overall shapes, change in any significant way.}
\label{f_dpa}
\end{figure}

\subsection{No inner hole}
\label{s_nohole}

In the previous section we have seen that our point source 
calculations yield similar behaviour in both \%Pol and PA as the analytical 
work of WBF. However, for an undisrupted disk that reaches the stellar surface, the 
behaviour is markedly different as can be seen in Fig.~\ref{f_dpa} -- for a range of 
inclination angles. The largest PA rotation seen is for an inclination of 15\degree; the 
subsequent inclinations plotted are 30\degree, 45\degree, 60\degree, and 75\degree. 
The level of continuum polarization 
is, as expected, higher for higher inclinations -- typically it reaches a few percent (but not shown, 
as neither the widths, nor the overall shapes, change in any significant way).

\begin{figure}
\centerline{\psfig{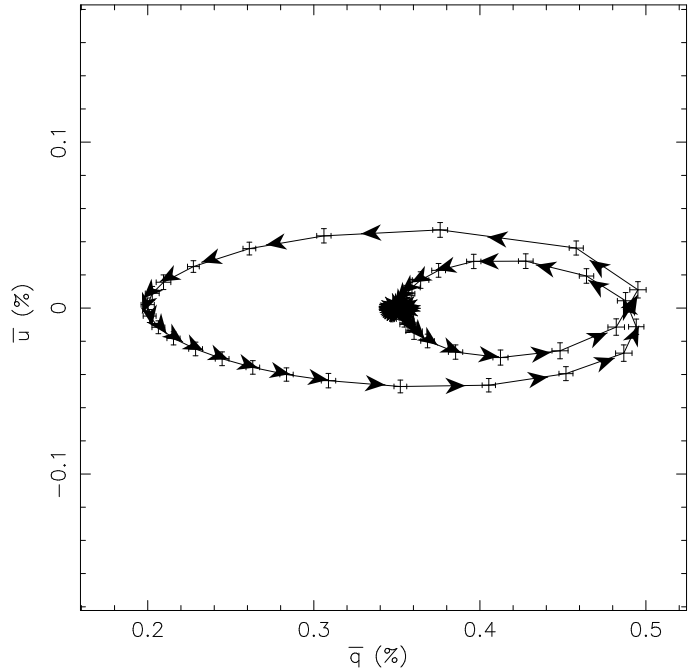}}
\caption{$QU$-plane representation of a double PA rotation. Actually, the normalized 
Stokes parameters $u = U/I$ and $q = Q/I$ are plotted. 
The figure represent the case of an undisrupted disk, at an inclination of 15\degree. 
The continuum points are situated at the $(Q,U)$ position of (0.35,0), and the direction of 
the arrows denotes the increase in wavelength across the line. There are 2 ellipses visible, 
representative of the double PA rotations in Fig.~\ref{f_dpa}.}
\label{f_qu}
\end{figure}

Most notable are the {\it double} PA rotations. This turns out to be 
the result of the finite size of the star.
One might perhaps expect that, even for a finite-sized transparent star 
(and single scattering only), there should not be any PA rotation, as there is 
no obvious front-back asymmetry in the disk geometry system. For the case of 
the central point source (as described in Sect.~\ref{s_test} above), we found 
that only a front-back asymmetry, such as 
occultation or multiple scattering, allows the Stokes $U$ parameter to rotate, 
since the scatterers have no radial (but only tangential) velocity with respect to 
the photon source, and the observed velocity field is necessarily front-back symmetric. 
However, once one adapts a finite-sized photon source, one effectively 
creates photon sources that are off-centre. This introduces a whole range of angles 
of incoming photons for every scatterer. Since most scatterers now have a small 
radial motion with respect to the source, the observed velocity field is 
no longer symmetric. Instead, it results from a combination of {\it two} 
relative velocities: there will be a Doppler shift associated with the 
source-scatterer difference as well as that for the scatterer-observer. For any given 
velocity bin on, say, the approaching side of the disk, 
there is now a front-back asymmetry due to the small scatterers' radial motion 
with respect to the off-centre photon source that may result in a PA rotation. But 
since the geometry between the front and back side of the disk itself is the same, 
this implies that {\it within} the blue-shifted part of the line,  
there {\it has} to be an additional turning point, which therefore results in the 
double PA rotation observed in the graphs of Fig.~\ref{f_dpa}. So, it is essentially 
the finite-sized star interacting with the disk's rotational velocity field, which 
re-sorts the scattered line emission, to cause the double PA rotations. 
Finally, note that the resulting double PA rotation, translates into 
{\it two} loops in $QU$ space, as seen in Fig.~\ref{f_qu}. 

\subsection{Varying the inner hole size}
\label{s_0spot}

\begin{figure}
\centerline{\psfig{file=1463fig4.ps, width = 9 cm}}
\caption{Varying the size of the disk inner edge: \Rin\ = 1,2,3 and 5\Rstar\ for an inclination 
of 45\degree. The dotted line is for the model \Rin\ = 1 \Rstar\ (the undisrupted disk), which 
has the largest PA amplitude.}
\label{f_vary1}
\end{figure}

\begin{figure}
\centerline{\psfig{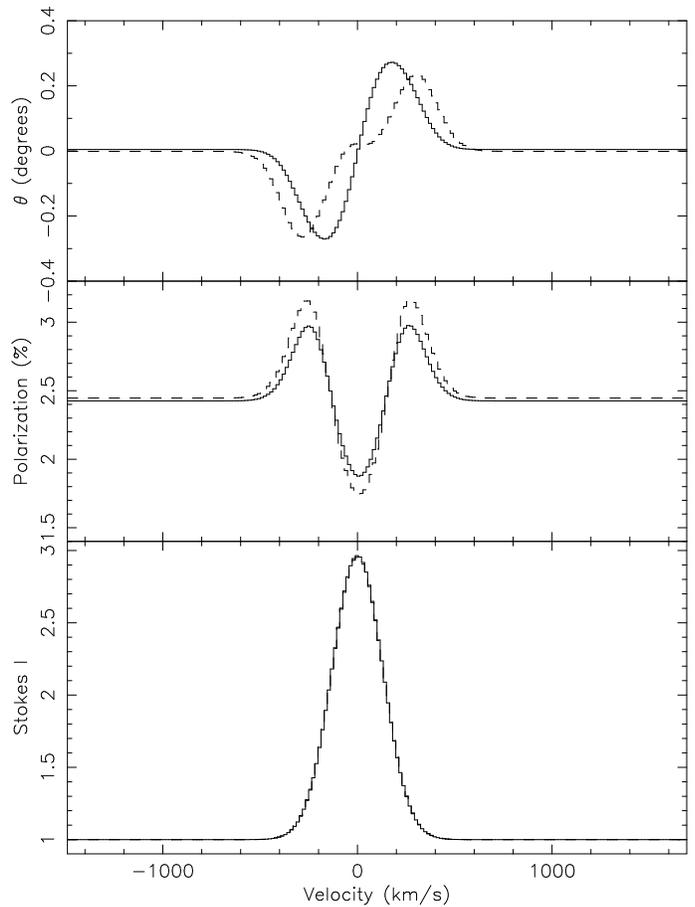}}
\caption{Similar as Fig~\ref{f_vary1}, but now with the PA axis blown-up, and only with the 
cases of \Rin\ = 3 (dashed) and 5\Rstar (solid) repeated.}
\label{f_vary2}
\end{figure}

\begin{figure*}
\mbox{
\epsfxsize=0.5\textwidth\epsfbox{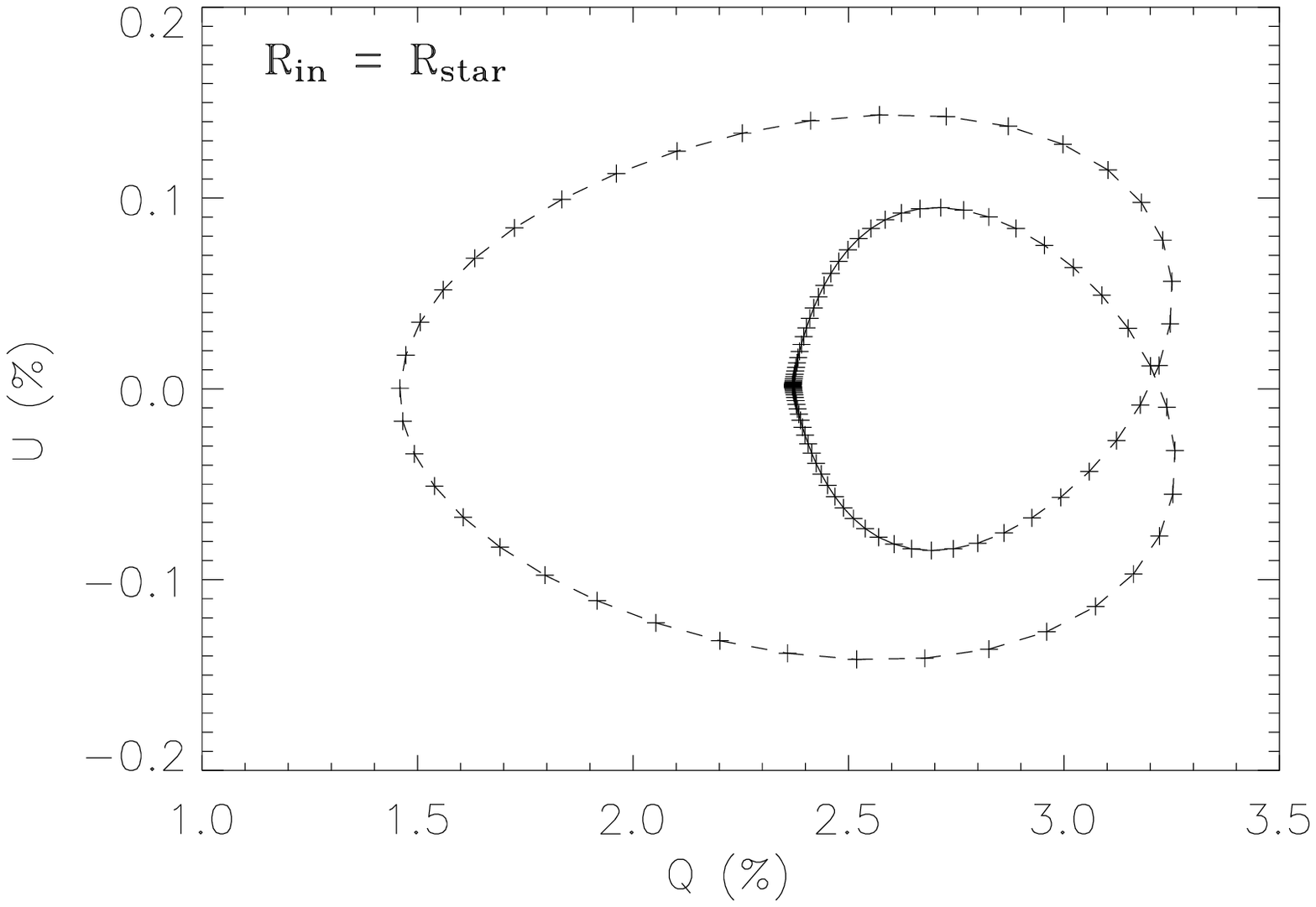}
\epsfxsize=0.5\textwidth\epsfbox{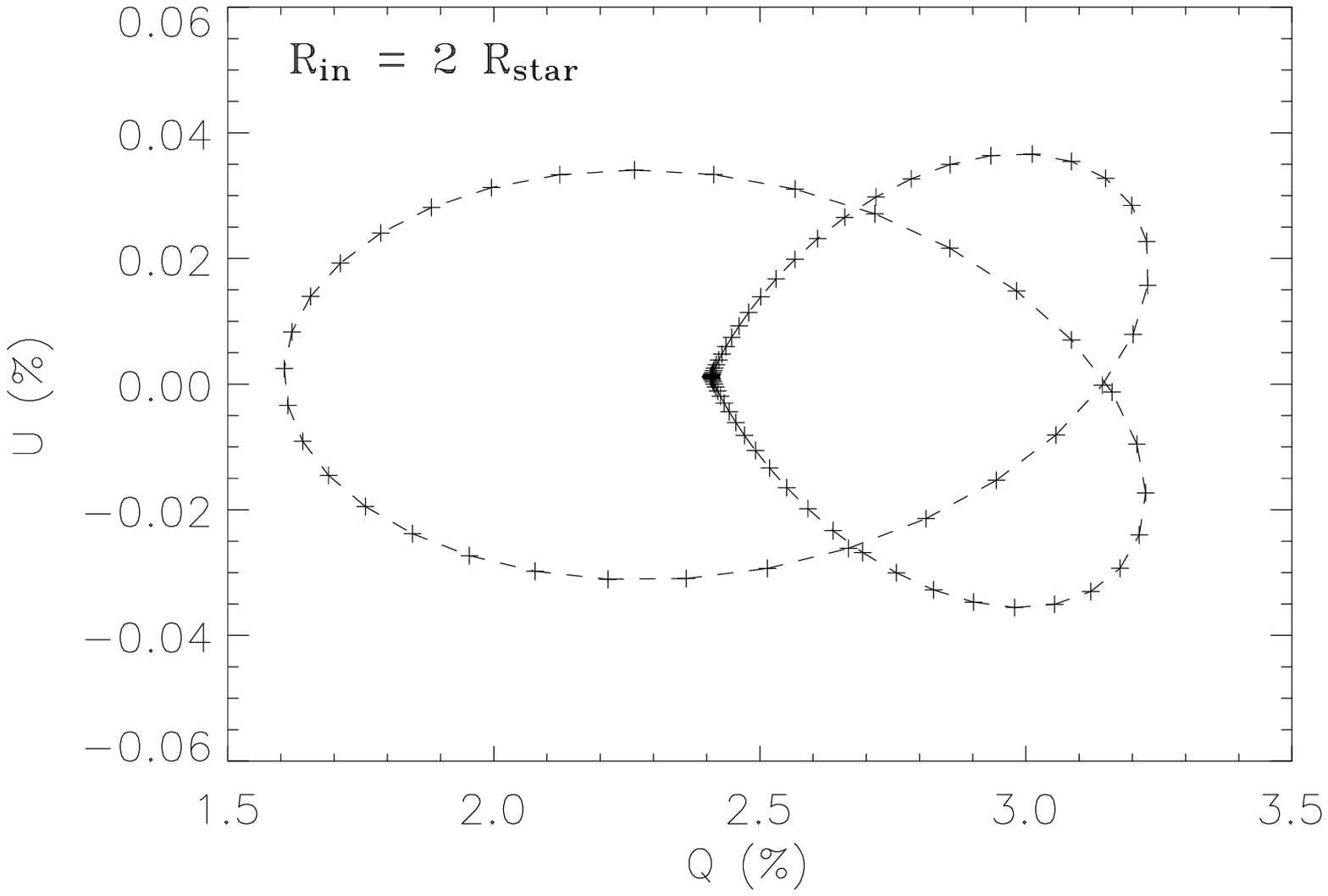}
}
\mbox{
\epsfxsize=0.5\textwidth\epsfbox{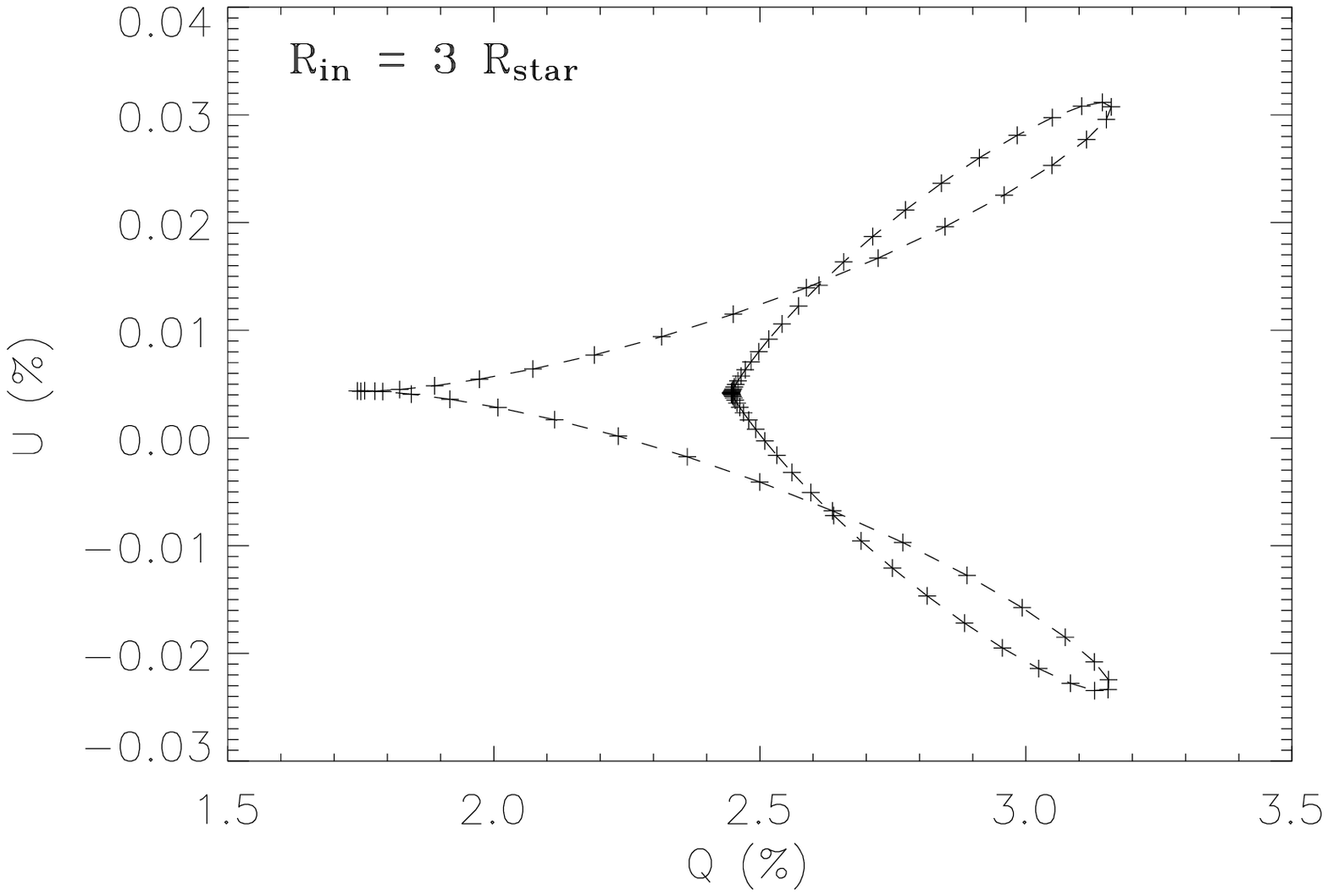}
\epsfxsize=0.5\textwidth\epsfbox{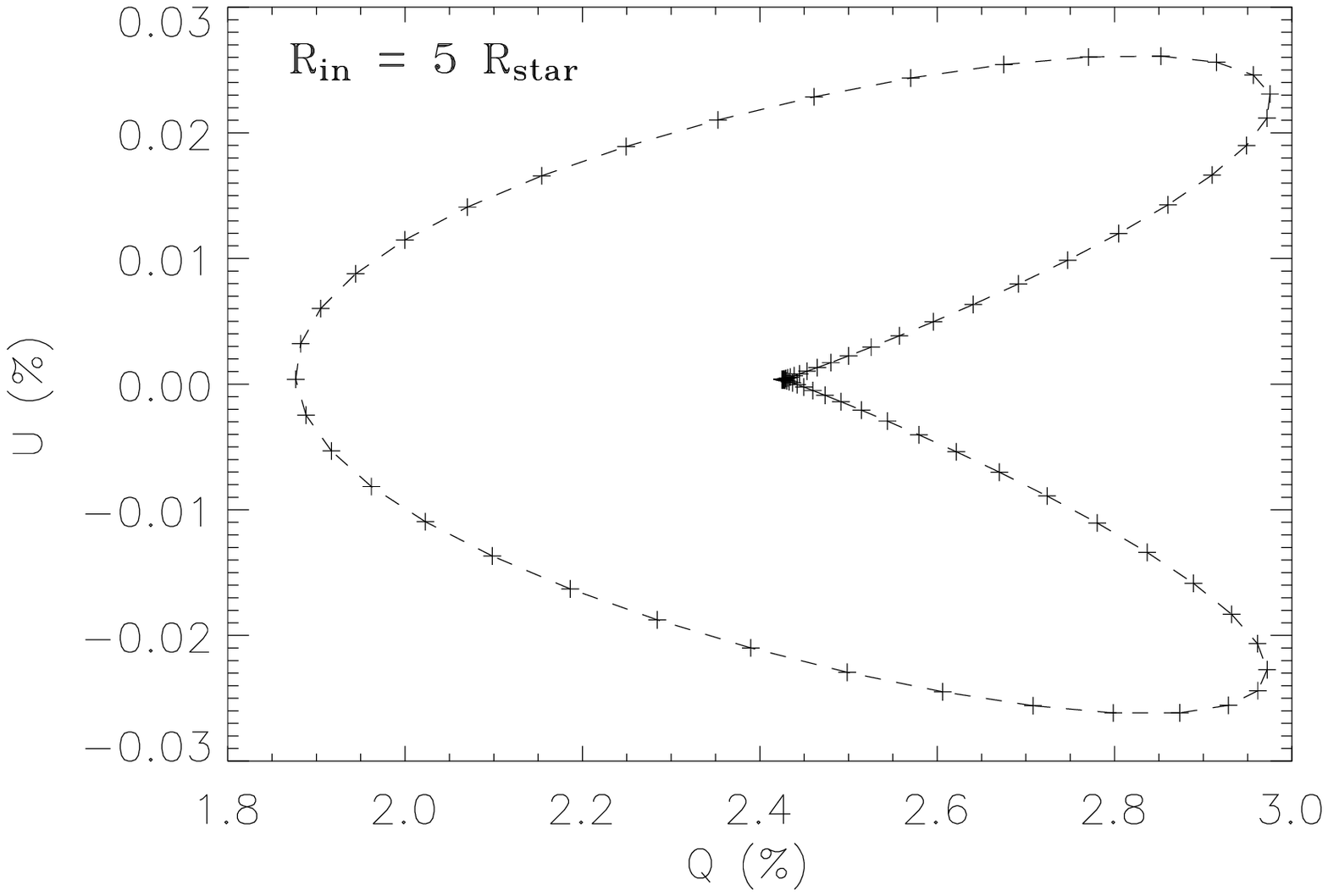}
}
\caption{$QU$ plane representation of the triplots of Fig.~\ref{f_vary2}. 
Actually, the normalized Stokes parameters $u = U/I$ and $q = Q/I$ are plotted again. 
The transition from \rin\ = 1,2,3 to 5 \rstar\ is accompanied with a $QU$ transition 
with the double loop for the undisrupted case, with \rin\ = 1 \rstar, through a ``butterfly'' 
shape in the diagram for 3 \rstar, to a single $QU$ loop for 5 \rstar. 
The continuum points are situated at a $(Q,U)$ position of approximately (2.5,0), and the 
direction of increasing wavelength, as most easily identified in the last plot, is clockwise.
Note that at $i = 45$\degree\ these shapes are flatter (narrower in $U$) than at $i = 15$\degree, 
cf. Fig.~\ref{f_qu}.}
\label{f_qu_vary}
\end{figure*}

We now introduce an inner hole (for a star with a core radius \Rstar), and 
study the effect of the size of an inner hole on both the shapes and the amplitudes 
of the polarization signatures. 
We start off with an undisrupted disk and subsequently increase the disk inner edge 
\Rin\ to 2, 3, and 5 \Rstar\  -- for an intermediate inclination angle of 45\degree. 
This sequence is graphically shown in Figs.~\ref{f_vary1} and~\ref{f_vary2}. Figure ~\ref{f_vary1} 
shows the full sequence, while Fig.~\ref{f_vary2} 
is a blow-up of the y-axis to better show the PA behaviour for the models with 
inner edges of 3 and 5 \Rstar.
For all models, we keep the same inner disk density, so that the continuum polarization 
properties at the inner edge do not change. 
The sequence of Fig.~\ref{f_vary1} shows that significant differences occur in the 
behaviour of the position angle. For a small inner hole (or no hole at all), the PA 
amplitude is relatively large -- up to about 2\degree\ for this particular inclination -- 
and the qualitative behaviour is that of a double PA rotation, as explained above. 
For larger hole sizes, the PA behaviour is first transitional (seen for \Rin\ $=$ 
3 \Rstar\ as the dashed line in Fig.~\ref{f_vary2}), 
and from 5 \Rstar\ onwards stabilizes to a clear {\it single} S-shaped rotation (seen as 
the solid line in Fig.~\ref{f_vary2}), as is typical for point sources.
This transitional PA behaviour is also visible in the $QU$ plane (Fig.~\ref{f_qu_vary}). 
The transition from \rin\ = 1,2,3 \rstar\ to 5 \rstar\ is accompanied with a $QU$ transition 
from a double loop at 1 \rstar, through a ``butterfly'' shape in the diagram for 3 \rstar, 
to a single (stretched-out) $QU$ loop in the case of 5 \rstar. 
Note that the differences in the level of the polarization percentage are only modest, as the inner disk 
density is fixed, and the subtended angle does not change significantly.  
As far as the amplitude of the PA change is concerned, it first decreases, but  
remains constant (at approximately 0.2\degree\ for this particular inclination angle), as the 
subtended angle no longer changes on increasing the inner hole size. 

The qualitative behaviour of the PA is not the same for all viewing angles, as was evident 
in Fig.~\ref{f_dpa} for the undisrupted disk. 
Instead of showing complete sequences for all possible inclinations, we 
first construct a table indicating the presence of 
{\it single} (``S'') and {\it double} (``D'') PA rotations.
Note that in between these ``S'' and ``D'' PA rotations, one may  
observe transitional behaviour, that we have termed ``t''. 

\begin{table*}
\caption{The boundaries between the {\it single} ``S'' and {\it double} ``D'' 
PA rotation -- for a range of inclination angles. 
By convention 90\degree\ is edge-on, 0\degree\ is pole-on. In between these ``S'' and ``D'' PA 
rotations, one finds transitional (``t'') behaviour.}
\label{t_boundary}
\begin{tabular}{lccccccccccccccc}
\hline
Incl(\degree) & 15 & 20 & 25 & 30 & 35 & 40 & 45 & 50 & 55 & 60 & 65 & 70 & 75 & 80 & 85 \\
\hline 
\Rin(\rstar)  &   &   &   &   &   &   &   &   &   &   &   &   &   &   &  \\
\vspace{0.2mm}
1             & D & D & D & D & D & D & D & D & D & D & D & D & D & S & S \\ 
2             & D & D & D & D & D & D & D & D & D & t & t & S & S & S & S \\ 
3             & D & D & D & t & t & t & t & t & t & S & S & S & S & S & S \\  
4             & t & t & t & t & S & S & S & S & S & S & S & S & S & S & S \\
5             & S & S & S & S & S & S & S & S & S & S & S & S & S & S & S \\ 
\hline
\end{tabular}
\end{table*}

\begin{figure}
\centerline{\epsfig{file=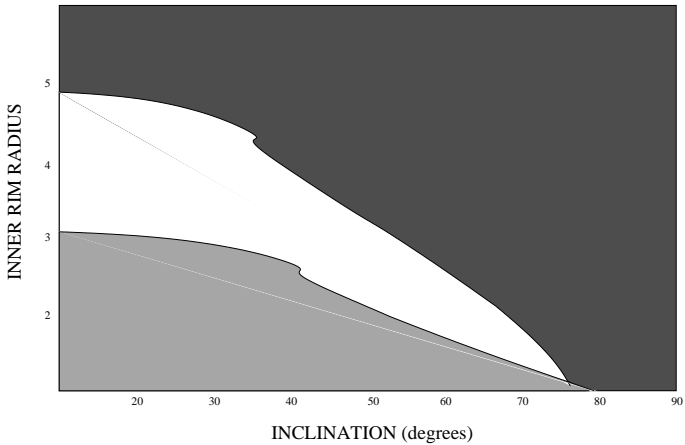, width = 9 cm}}
\caption{The boundaries between the {\it double} (grey shaded) and {\it single} (black)  
PA rotations for a range of inclination angles. 
By convention, 90\degree\ is edge-on, 0\degree\ is pole-on. In between the single and double 
PA rotations, one finds transitional (white) behaviour. Note that the kinks in the boundaries 
are artifacts of the plotting technique.}
\label{f_boundary}
\end{figure}

Table 1 indicates that for a known inclination the behaviour of the PA rotations 
may constrain the size of the inner hole. 
For clarity, we have transformed the results from Table 1 into Fig.~\ref{f_boundary}.
The figure shows the areas where {\it single} (black shaded), {\it double} (grey), and 
{\it transitional} (white) PA behaviour occur. It indicates that for roughly pole-on 
disks (with $i$ $\la$ 40\degree), double PA rotations appear for \Rin\ $\la$ 3\Rstar, 
while for intermediate inclinations a double PA rotation is only present for an  
undisrupted disk, or a smaller inner hole, \Rin\ $<$ 2\Rstar. 
Finally, for high inclination (edge-on) models ($i$ $\ge$ 85 \degree) all 
models yield a single PA rotation. 

\begin{figure}
\centerline{\psfig{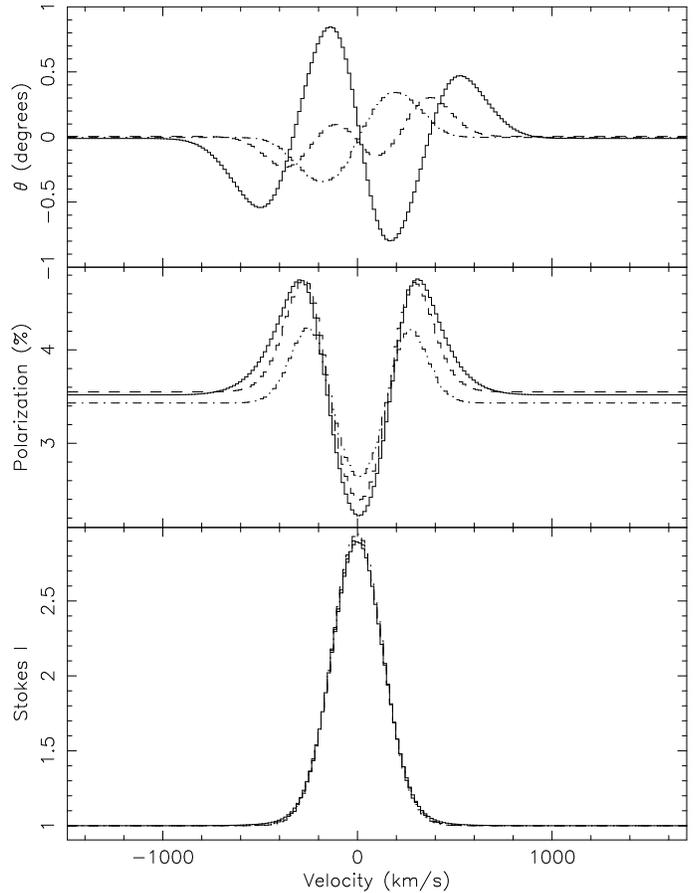}}
\caption{Similar to Figs.~\ref{f_vary1} and~\ref{f_vary2}, but for an inclination of 60\degree.
The solid line is for the undisrupted disk of 1\Rstar, the dashed line is for a disk inner 
rim at 2\Rstar, while the dashed-dotted line is for a disk truncation radius of 5\Rstar.}
\label{f_60}
\end{figure}

\begin{figure}
\centerline{\psfig{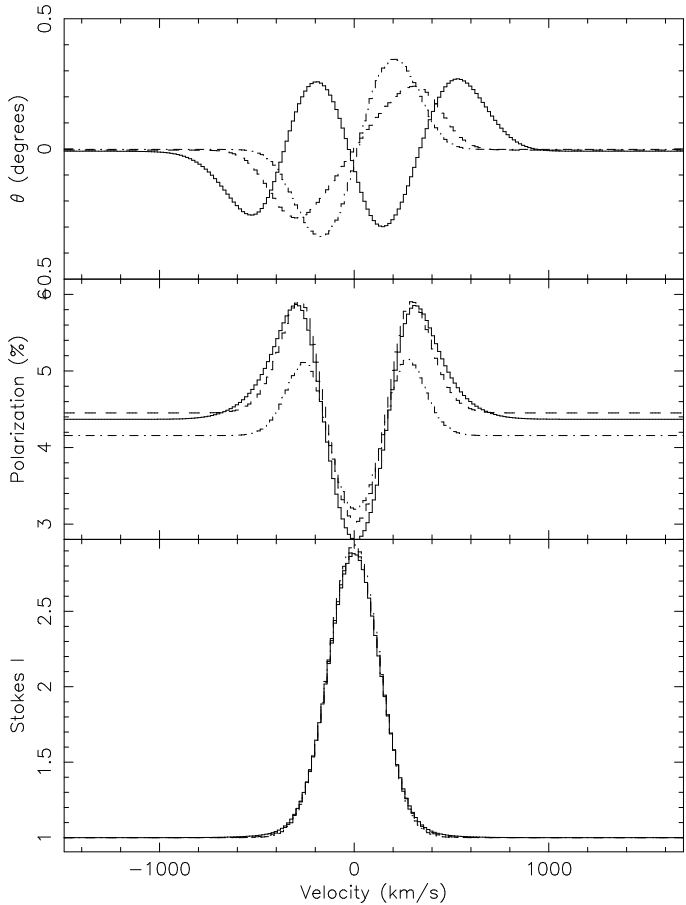}}
\caption{Same as in Fig.~\ref{f_60}, but for an inclination of 75\degree.}
\label{f_75}
\end{figure}

Table~1 and Fig.~\ref{f_boundary} gloss over what is in reality the complex 
nature of the PA behaviour in the transitional region. 
To demonstrate this, we show graphical sequences 
for inclinations of 60\degree\ and 75\degree\ in 
Figs.~\ref{f_60} and~\ref{f_75}.
The solid line in Fig.~\ref{f_60}, for an inclination of 60\degree, shows the 
double PA rotation for \Rin\ $=$ 1 \Rstar, but these models are found 
transiting for \Rin\ $=$ 2 \Rstar, as indicated by the dashed line. At larger hole sizes, the 
dashed-dotted line in Fig.~\ref{f_60} shows the single PA rotation typical 
for a point-like source. Increasing the inclination further, to 75\degree\ (Fig.~\ref{f_75}) 
demonstrates that the PA amplitudes are lower than for 60\degree, as expected. 
Although the single PA rotation is here already reached at \Rin\ $=$ 2\Rstar, the shape of the 
PA flip changes as hole size is further increased (cf. the dashed line of 2 \Rstar\ and the dashed-dotted 
line, at 5 \Rstar). In particular, the \Rin\ $=$ 2 \Rstar\ case shows a more gradual 
change across the line profile, whereas the larger hole sizes show a more rapid change. 
This demonstrates that for the more edge-on disks, where the transition from double PA to single PA
rotation remain absent according to Table~1, 
the shape of the single PA flip can still have similar diagnostic value for 
determining the inner hole size (provided that the data exhibit sufficient
signal-to-noise, see Sect.~\ref{s_discuss}).

\begin{figure}
\centerline{\epsfig{file=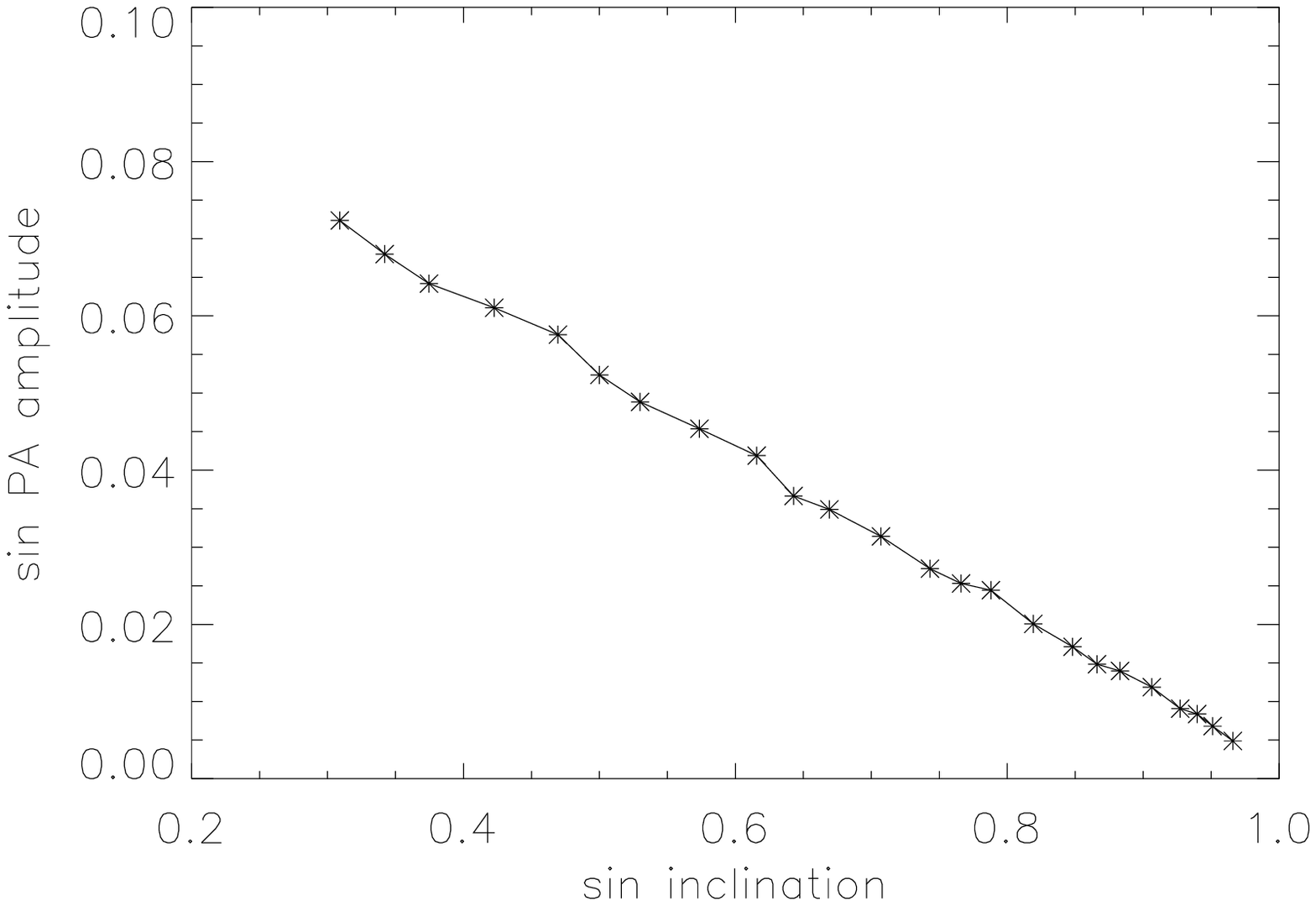, width = 9 cm}}
\caption{The amplitude of the double PA rotation, $\sin$ $\Delta$ PA, as a function 
of disk inclination, $\sin$ $i$, for the theta disk. The solid line connects the 
individual data points.}
\label{f_pa}
\end{figure} 

We have seen that the amplitude of the PA rotations is a rather strong function 
of inclination, and we have plotted these amplitudes (sin PA) for the case of no hole 
against the disk inclination (sin $i$). This inclination dependence of the PA amplitude is 
shown in Fig.~\ref{f_pa}. 
The results for the theta disk  
were found for the parameters: \taud\ $=$ 1, 
and \optheta\ $=$ 10\degree. Note 
that we have ran the models for different choices of the parameters \taud\ and 
\optheta\ (in the ranges of \taud\ $=$ 1 -- 10, and \optheta\ $=$ 3 -- 10\degree), but we found 
no significant changes to either the amplitude of the PA 
rotations, or their shape characteristics. 
This is not surprising, since the PA behaviour is a geometric and kinematic effect, and 
does not depend on the exact level of continuum polarization, although the parameters chosen 
do play a dominant role in setting the degree of linear polarization, according 
to the Brown \& McLean (1977) results, such as Eq.~(\ref{eq_brown}), and our own computations 
in Sect.~\ref{s_test}.


\section{Discussion}
\label{s_discuss}

We have computed polarimetric line profiles for scattering off rotating 
disks with a Monte Carlo model, but we have not yet discussed competing effects and other 
disk geometries, nor have we attached any preference to a particular type of 
scattering particle.

First of all, it should be noted that in spectropolarimetric observations 
there is always the addition of interstellar polarization to that intrinsic to the source 
of interest. This implies that the observed continuum PA is generally not equal to zero and 
that the level of continuum polarization will be affected by a foreground contribution
due to interstellar grains.
Nonetheless, the {\it differential} effect between line and continuum will not be affected by 
foreground polarization, in that the shapes and sizes of the loops in the $QU$ plane 
remain exactly the same (they are just shifted within the plane).

Second, there is the issue of unpolarized line emission. For classical Be stars the emission 
line is generally believed to be formed in the circumstellar environment, rather than at 
the stellar surface, as assumed in our models. 
We therefore note that there are two potential polarisation effects, which could sometimes
be at work at the same time in some astrophysical objects.  
The two effects being the depolarizations due to the unpolarized line emission from a large volume, 
and the intrinsic line polarization due to a compact photon source. 
We note that the position angle rotations due to intrinsic line polarization as predicted 
in this paper should be easily distinguishable from depolarizing effects, because 
of their different characteristics in the $QU$ plane (see Vink et al. 2002).

We also note that we have assumed idealized disk geometries. For instance, 
for the application of our polarization models to the disks around 
pre-main sequence stars, one may need to consider a more sophisticated ``flaring'' disk geometry,  
since infrared spectral energy distribution (SED) modelling has indicated a preference 
for such flaring disks (Kenyon \& Hartmann 1987; Chiang \& Goldreich 1997) rather than disks with a constant
opening angle. These flaring disks may even posses puffed-up inner rims (e.g. Dullemond et al. 2002). 
As our main purpose here is to highlight the unexpected effect of the disk inner radius on the line polarization,
we have not modelled a flaring disk here. 
We are aware that a flaring disk could intercept light at larger distances from the star, which would 
likely contribute to the percentage polarization of the continuum. However, the differences between 
a flared and a theta disk are expected to be only subtle as far as the predicted
polarization changes across spectral line profiles are concerned.

The issue of the polarizing agent has not yet been addressed. For hot stars, such as 
classical Be stars, the polarizing agent is generally believed to be scattering electrons (although 
hydrogen continuum opacity does also have an effect on the broad band 
spectropolarimetry, especially at ionization edges). These electrons are known to be 
able to smear out line polarimetric profiles due to their potentially large thermal motions 
compared to the bulk motions of the stellar envelope (see Wood \& Brown 1994).
Therefore, some of the polarization rotation structure across emission lines  
would be diminished for hot stars by this thermal broadening effect. This would especially be 
true for lower sensitivity measurements. To date, line polarimetry with 4m-class telescopes 
is usually only performed at the S/N level of 1000, and the accuracies are therefore 
of the order of about 0.1\% (see Tinbergen \& Rutten 1997). Note that demands 
on differential measurements (i.e. across a spectral line) are less severe than absolute 
ones. Already some of the published data from the Anglo Australian Telescope (AAT) and the 
William Herschell Telescope (WHT) have performed better than at the 0.1\% level (see for instance 
the data on the O star Zeta Pup by Harries \& Howarth 1996).
Nonetheless, in the current era of 8m-class, and in the upcoming era of 100m-class telescopes, 
larger photon collecting areas will make routine high precision spectropolarimetry feasible, 
such that even in the presence of thermal broadening, subtle changes in the 
polarization and PA may nonetheless be measurable.

For cooler stars, dust may be the principle polarigenic agent.
As mentioned before, although the matrices 
for Mie and Rayleigh scattering, as used in our study, are different, both favour forward and 
backward scattering, so the differences between our predictions and those for Mie scattering are 
expected to be only qualitative. Note that dust grains are not expected to have large thermal velocities that would 
result in significant line broadening.
Another potential opacity source for which smearing is not expected to be significant is that 
of neutral hydrogen, which has for instance been invoked in the case of symbiotic stars
(Schmid 1995, Harries \& Howarth 1997, Lee \& Lee 1997). 
This may imply that the results of this study are of greater relevance to
cooler stellar objects. Obvious cool star candidates are post-AGB stars (where disks are probably 
present given the bipolarity of many post-AGB nebulae), 
as well as in the lower mass PMS stars, such as T Tauri stars, where hydrogen is expected 
to be neutral, and dust may survive rather close to the stellar surface. 

We have mentioned the high demand on sensitivity, but another aspect is that 
of spectral resolution. 
Currently, the resolution that can be achieved on 
common-user optical instruments is limited to 
$R$ $\la$ 10 000 (e.g. Schulte-Ladbeck et al. 1994, Oudmaijer et al. 1998, Vink et al. 2003).
To check whether the PA changes as predicted in this paper would be resolvable with these
instruments, we have degraded our model spectra to $R$ $=$ 10 000, 
and found that the instrumental resolution should not wipe out the predicted $QU$ profiles.
The greater limitation is indeed sensitivity. This is because in the handling of  
spectropolarimetric data 
a degeneracy between sensitivity and spectral resolution is usually maintained. Currently, one
generally needs to rebin pixels across a spectral line, 
to gain the signal needed to achieve the required polarimetric accuracy. 
We conclude that, whilst there is already plenty of evidence of single PA rotations in 
e.g. Herbig Ae stars (Vink et al. 2002), that are entirely consistent with the idea of disrupted disks 
in these objects, the sensitivity is typically too poor to allow for quantitative comparisons 
between models and data. In any case, no instance of a double PA flip has yet been positively 
identified. However, this must be seen as absence of evidence rather than evidence of absence 
until appreciably greater sensitivity (with S/N $>>$ 1000) becomes routine.
Interestingly, recent data on the bright T~Tauri star GW~Ori show the presence of a gradual PA change 
across \ha\ (Vink et al. 2004, in preparation), which may be indicating the presence of a relatively small 
inner hole, \Rin\ $\le$ 2 \Rstar, for an inclined disk ($i$ $\simeq$ 75\degree), or a somewhat 
more pole-on disk, with a larger inner hole. 

Note that most of the PA rotation amplitudes derived in this paper (dropping to only a 
few tenths of a degree in some of the models of inclined disks), correspond to 
changes in Stokes $U$ of only 0.05 \% or less throughout the spectral line. 
Upon occasion, this is measurable with 
today's instrumentation, provided the integration times are long enough. 
This is well worth the effort, since, as we have shown here, line polarimetry
can uniquely obtain combined constraints on disk inclination and inner hole radius.

\section{Summary}
\label{s_sum}

In this paper we have predicted polarimetric line profiles for scattering 
off rotating disks using a Monte Carlo model that removes geometric approximations required 
in earlier work. We have found the following new results:

\begin{itemize}

\item{} Multiple scattering can play an analogous front-back asymmetry to disk scattering 
as the effect of stellar occultation. Both 
effects result in an S-shaped rotation of the position angle.

\item{} There is a marked difference between scattering of line emission 
by a disk that reaches the stellar surface, and a disk with an inner hole. 
For the case with an inner hole, we find the familiar 
single PA rotations, while for an undisrupted disk, we find double rotations.

\item{} The effect of double PA rotations is due to the finite-sized star 
interacting with the disk's rotational velocity 
field, which re-sorts the scattered line emission.

\item{} Since a gradual increase of the hole size transforms the double rotations smoothly back 
into single ones -- as the line emission object approaches that of a point source -- our 
models demonstrate the diagnostic potential of line polarimetry in determining not 
only the disk inclination, but also the size of the disk inner hole.

\end{itemize}

Our computations are simple and idealized: although they are performed for 
stellar parameters most appropriate to Herbig Ae stars, they may  
also be relevant to a range of other centrally-condensed objects embedded in disks. 
These may include a variety of objects as diverse as pre-main sequence stars to 
active galactic nuclei.

\begin{acknowledgements}
We would like to thank Rene Oudmaijer, Stuart Sim, and an anonymous referee for providing 
insightful comments. JSV is funded by the 
Particle Physics and Astronomy Research Council of the United Kingdom.  
\end{acknowledgements}

\end{document}